\documentclass{article}
\usepackage{PRIMEarxiv}
\usepackage{amsmath}
\usepackage{siunitx}
\usepackage[T1]{fontenc}    
\usepackage{hyperref}       
\usepackage{url}            
\usepackage{booktabs}       
\usepackage{amsfonts}       
\usepackage{nicefrac}       
\usepackage{microtype}      
\usepackage{lipsum}
\usepackage{fancyhdr}       
\usepackage{graphicx}       
\usepackage{xcolor}
\usepackage{caption}

\newcommand{\inhabitants}{P}
\newcommand{\elongation}{E}
\newcommand{\unusedArea}{S}
\newcommand{\height}{F}
\newcommand{\profile}{W}
\newcommand{\perresident}{A}

\newcommand{\perBuilding}{B}

\newcommand{\network}{N}
\newcommand{\connections}{C}
\newcommand{\vertices}{V}
\newcommand{\vertex}{v}
\newcommand{\connection}{c}

\newcommand{\heightsOfBuildings}{H}

\title{What is the best shape of a city}

\author{
Tobias Batik \\ 
Complexity Science Hub \\ Metternichgasse 8, 1030 Wien \\ 1080 Vienna, Austria \\
\And
Guillermo Prieto-Viertel \\ Complexity Science Hub \\  Metternichgasse 8, 1030 Wien \\ 1080 Vienna, Austria \\
\And
Jiaqi Liang \\ Complexity Science Hub \\ Metternichgasse 8, 1030 Wien \\ 1080 Vienna, Austria \\
\And
Liuhuaying Yang \\ Complexity Science Hub \\ Metternichgasse 8, 1030 Wien \\ 1080 Vienna, Austria \\
\And
Dániel Kondor \\ Complexity Science Hub \\ Metternichgasse 8, 1030 Wien \\ 1080 Vienna, Austria \\
\And
  Rafael Prieto-Curiel\\
  Complexity Science Hub \\ Metternichgasse 8, 1030 Wien \\ 1080 Vienna, Austria \\
  \texttt{prieto-curiel@csh.ac.at}
}

\begin{document}
\maketitle

\section{Abstract}

Urban form plays a crucial role in shaping transportation patterns, accessibility, energy consumption, and more. Our study examines the relationship between urban form and transportation energy use by developing a parametric model that simulates city structures and their impact on travel distances. We explore various urban morphologies, including sprawling, elongated, compact, and vertically concentrated cities, and consider five urban profiles: ``needle'', ``pyramid'', ``pancake'', ``bowl'' and ``ring''. We designed an interactive visualisation and calculator that enables the analysis of these effects, providing insights into the impact of various urban configurations. Our model quantifies the average commuting distances associated with these forms, demonstrating that compact and centrally dense cities minimise the total travel distance in cities.


\section{Introduction}

{
Urban form, understood as the physical forms and spatial arrangements of human settlements, plays a central role in shaping some of the most pressing sustainability challenges \cite{zhang2023spatial}. Among these, the influence of urban form on transportation systems and their associated energy use is critical, given that urban areas account for approximately 75\% of global energy use and 70\% of greenhouse gas emissions \cite{IEA2018}. The distances people must travel daily are mainly determined by the structure of a city and represent a key driver of urban energy consumption and related emissions \cite{kaza2020urban}. Besides transport, the shape of a city and how it grows also drives other urban outcomes, such as the impact of heat waves \cite{MASHHOODI2024105690}, land cover change and biodiversity loss \cite{seto2010new}. At the human scale, it profoundly affects the lived experiences of city residents, influencing health outcomes, daily accessibility, and social cohesion \cite{Miles2012, Bramley2009, Frank2001, Lenzi2023}. Yet the ways in which cities grow can result in urban forms that undermine both sustainability and quality of life \cite{duque2019urban}. Unplanned city growth often leads to fragmented urban landscapes and social segregation, where poor connectivity and uneven access to essential services, such as healthcare, education, and sanitation, disproportionately impact the most vulnerable \cite{brelsford2018cities, corburn2016informal, mihigo2022effects, legeby2010urban, WaterPrietoBorjaArxiv}. Even planned cities often fall short of their ambitions, as the inflexibility of the plans, flawed execution, or a disconnect between the assumptions made in planning and residents' real needs can all contribute to these shortcomings \cite{Jacobs1961, Watson2009, Faria2013}. Studying urban form is, therefore, essential to planning and building more equitable, resilient, and sustainable cities.
}

{
Specific urban form attributes, such as density, spatial configuration, and city shape, translate these spatial relationships into concrete mobility behaviours and emission patterns. Elongated and sprawling cities, characterised by low-density development and extended urban footprints, tend to increase average travel distances, foster car dependency, and elevate transportation-related energy use and emissions \cite{osorio2017understanding, monteiro2024challenges, prieto2023scaling}. This spatial fragmentation not only reduces accessibility but also contributes to traffic congestion, time losses, and environmental degradation \cite{resch2016impact, gossling2020why}. In contrast, compact cities concentrate amenities and services, shorten trip distances, and support more sustainable mobility options such as walking, cycling, and public transit \cite{Foraboschi2014, batty2008scaling, rossi-hansberg2009firm}.
}

{
Beyond the horizontal extent, the vertical shape of cities, defined by building height and its spatial profile, plays a critical role in shaping transport energy outcomes. ``Pyramid'' cities, characterised by dense, high-rise cores that descend into lower peripheries, are often associated with shorter per capita travel distances and greater transit accessibility \cite{zhou2022satellite, lall2021pancakes}. In contrast, flat or ``pancake'' cities, with uniformly low buildings spread evenly across the urban area, tend to promote longer trips, higher private vehicle use, and increased infrastructure demands \cite{zhou2022satellite, lall2021pancakes}. While taller buildings require more energy for operations like heating, cooling, and vertical circulation \cite{hamilton2017all, godoy2018energy}, their potential to reduce surface-to-volume ratios and enable compact land use often offsets these costs in dense environments \cite{schlapfer2015urban, bibri2020compact}. Moreover, the spatial distribution of building heights and open spaces significantly influences pedestrian accessibility, microclimates, and social cohesion, underscoring the multifaceted relationship between vertical urban form and urban sustainability \cite{schlapfer2015urban, Watson2009}. Taken together, these insights underscore the need for a multidimensional understanding of urban morphology to optimise accessibility, reduce energy demand, and enhance urban sustainability \cite{biljecki2022global}. Yet, despite this recognition, developing a unified analytical model that captures all relevant aspects of city form remains a significant challenge.
}

{
Previous research has attempted to approximate this complexity through simplified models of spatial structure. Some studies rely on analytical approximations to describe interactions among spatially distributed elements \cite{PointsInCitiesRibeiroRybski}, while others examine how morphological features like sprawl and elongation scale with city size \cite{prieto2023scaling}. Parallel efforts have introduced typological and classification frameworks to organise the multiple dimensions of city morphology based on spatial indicators \cite{biljecki2022global} and literature-based taxonomies \cite{zhang2023spatial, fleischmann2021measuring, burke2022geospatial}. While these approaches yield valuable insights, they fall short because they rely on approximations and often assume a uniform population distribution. For instance, the height of buildings in pyramid cities has been modelled as a function of city-wide elements, such as population and income, with wealthier and larger cities achieving a taller peak in the centre \cite{lall2021pancakes}. Yet, these many elements mask key real-world constraints and interactions among urban form characteristics that ultimately shape travel patterns. Consequently, it remains unclear how to integrate these partial insights into a comprehensive model of an optimal urban shape, one that minimises energy and mobility burdens.
}

{
We focus on how urban form influences travel distances within a city. While transportation is only one aspect of urban sustainability, its share in total energy use and carbon emissions is substantial. To quantify how different urban shapes influence travel behaviour, we develop a model that parametrises key dimensions of city form. Our framework generates synthetic layouts based on independent characteristics, including elongation, sprawl, average building height, population size, and vertical profile. These parameters enable us to replicate a variety of urban shapes, including configurations in which building heights vary with the distance to the city centre. For each configuration, we simulate commuting journeys and calculate the average travel distance as a proxy for mobility-related energy demand \cite{prieto2023scaling}. By systematically altering one parameter at a time, we isolate the spatial effects of urban form on mobility outcomes. Our model confirms that a compact and round city reduces travel distance, and that taller and more pyramidal shapes further reduce travel distances. While simplified by design, our model offers an intuitive and interactive way to explore how urban form influences mobility, making the complexity of these spatial relationships more comprehensible. An interactive version of the model, along with a calculator for various configurations based on simulated travel, is available at \url{https://vis.csh.ac.at/cities-morphology}, allowing users to explore different city profiles.  
}

\section{Results}

{
To understand how different urban forms affect commuting distances, we simulate cities with varying forms and measure travel within them. Our modelling strategy is driven by the need to quantify the different aspects of urban form in a context where experimental approaches are limited, interactions between numerous variables are complex, and commuting distances cannot be analytically measured. We model city layouts based on seven parameters: elongation, sprawl, average building height, population size, the number of people living in a building, floor area per resident, and city profile (Fig. \ref{fig:city-varibles}). The elongation of a city, denoted by $\elongation$, captures its horizontal shape and is defined as the ratio between the major and minor axes of the ellipse enclosing the city \cite{prieto2023scaling}. A value of $\elongation = 1$ corresponds to a circular city, while larger values indicate increasingly elongated or elliptical forms. The sprawl, denoted by $\unusedArea$, characterises the degree of dispersion of buildings within the urban area. It is measured as the average distance between neighbouring buildings and expressed in meters. In practical terms, this parameter represents the space allocated to streets, green areas, or unused land between structures \cite{ewing1997angelesstyle}. The average building height, denoted by $\height$, refers to the mean number of floors per building across the city, assuming uniform floor-to-floor height. In our model, an increase in $\height$ corresponds to a decrease in the footprint of individual buildings, maintaining the total residential volume per capita constant. The population size, denoted by $\inhabitants$, specifies the total number of residents in the simulated city and determines the aggregate demand for interior space. The number of residents per building, denoted by $\perBuilding$, sets the average number of individuals housed within a single building. Together with $\inhabitants$, this parameter defines the number of buildings required to accommodate the population. The floor area per resident, $\perresident$, specifies the average floor space used by a resident. Together with $\height$ and $\perBuilding$, it determines the footprint of individual buildings. Taken together, the parameters $\unusedArea$,  $\height$, $\inhabitants$, $\perBuilding$ and $\perresident$ determine the overall building footprint of the city.

\begin{figure}[h!]
    \centering
    \includegraphics[width=0.8\linewidth]{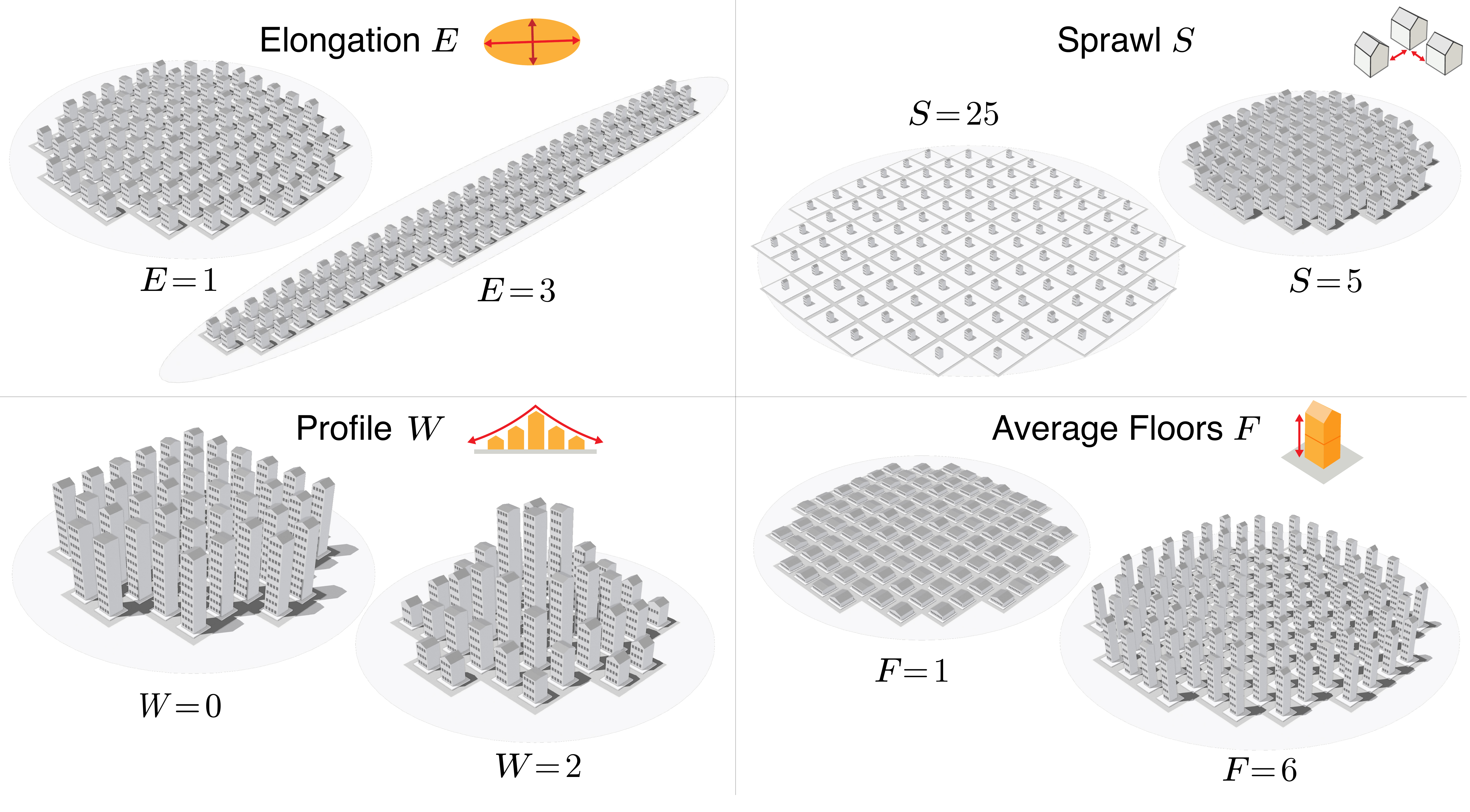}
    \caption{Effect of four key parameters on the urban form: elongation ($\elongation$), sprawl ($\unusedArea$), profile ($\profile$) on a city, and average building height ($\height$).}
    \label{fig:city-varibles}
\end{figure}
}

{
The profile of the city is denoted by $\profile$. It is a key parameter that defines if the building heights are evenly distributed in the city (forming a flat or pancake city), if there are taller buildings in the centre (forming a pyramid city), or if, inversely, there are shorter buildings in the centre and taller buildings by the periphery (forming a ``bowl'' city). For extreme values of the $\profile$, we obtain a ``needle'' for large values of $\profile \gg 0$; or a ``ring'' for negative values of $\profile \ll 0$ (Fig. \ref{fig:city-profiles}). 
\begin{figure}[h!]
    \centering
    \includegraphics[width=0.8\linewidth]{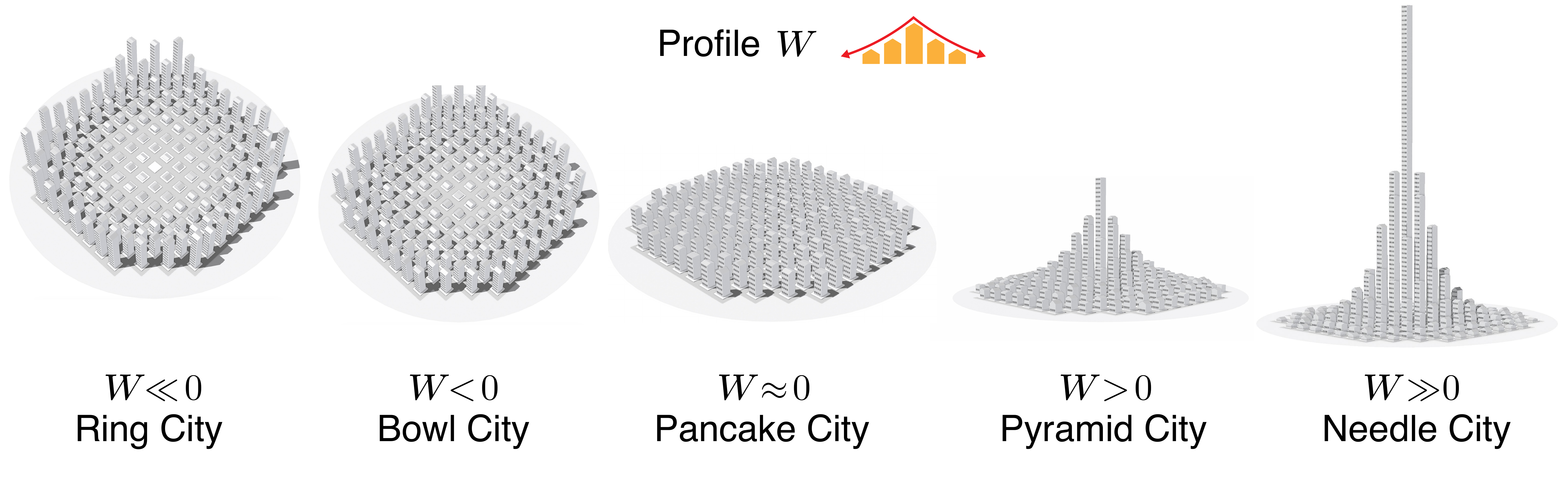}
    \caption{Different values of the profile of a city ($\profile$) result in different typologies, which we classify into five: a ring, a bowl, a pancake, a pyramid, and a needle city.}
    \label{fig:city-profiles}
\end{figure}
}

{
We quantify the impact of the parameters on cities' average travel distance by individually varying each value and generating a corresponding city layout. For each generated layout, we simulate the origin-destination journey of 100,000 individuals by sampling their starting and ending points as locations in that city. Sampling probabilities for each location are weighted by the number of people in the area, as determined by the city profile, to capture the trips that are more likely to occur. We measure the Manhattan distance between those pairs and compute the average travel distance. 
}

{
To anchor our simulation framework in an observed set of parameters, we empirically derive their values using data from Vienna. The total number of residents ($\inhabitants = 2$ million) and the average area per resident ($\perresident = 38\,\text{m}^2$) were obtained from official statistics on Vienna’s housing and population \cite{stadtwien2023gebaude}. The average number of persons per building ($\perBuilding = 11.4$) was calculated based on data from residential buildings and housing units \cite{stadtwien2023gebaude}. 
The sprawl was set to $\unusedArea = 21.2$ \si{\m} based on an estimated total streetwidth of 20 \si{\m} and an additional 10\% of an average building side ($\approx 1.2$). The average building height ($\height = 3\,$) was based on publicly available building height data \cite{stadtwien2024registerzaehlung}. The elongation ($\elongation = 1.5$) and profile ($\profile = 1.2$) were estimated to approximate the city's general layout and building height distribution. Using these parameters, our simulation produced an average travel distance of 9.3 \si{\km}. This result is well aligned with empirical data: 22\% of employees commuting between districts travel up to 4 \si{\km}, 62.5\% up to 9 \si{\km}, and about 85\% up to 14 \si{\km} \cite{statistikaustria2020abgestimmte}. Our simulated average of 9.3 \si{\km} falls within the central range of these observations, suggesting that the model captures the essential spatial dynamics of urban mobility in Vienna. Therefore, the model reasonably represents the urban structure and commuting behaviour in a simplified set of parameters.
}

\subsection{Impact of urban form on travel distances}

{
To isolate the effects of specific urban shapes, we vary four key shape parameters while holding other factors constant: sprawl ($\unusedArea$), elongation ($\elongation$), average building height ($\height$), and profile ($\profile$). Starting with the set of parameters that approximately resemble those of Vienna, we alter the values of the parameters individually by sampling values within each parameter range (Fig. \ref{fig:parametersVsTravelDistance}). We vary the elongation between values of $\elongation = 1$, which is observed when the city forms a circle, to $\elongation = 5$, observed for a very elongated form. The sprawl ranges from $\unusedArea = 0$\si{\m}, which corresponds to a compact city, to values of $\unusedArea = 60$ \si{\m}, indicating a highly sprawled city. The average number of floors ranges from $\height = 1$ for a flat city to $\height = 20$ for a more vertical layout. Since we maintain a constant number of residents per building, this variation also corresponds to adjusting the building's footprint accordingly. Finally, the profile of the city goes between $\profile = -20$, giving a city with a ring shape, to $\profile = 20$, which is a needle city.

\begin{figure}[h!]
    \centering
    \includegraphics[width=\linewidth]{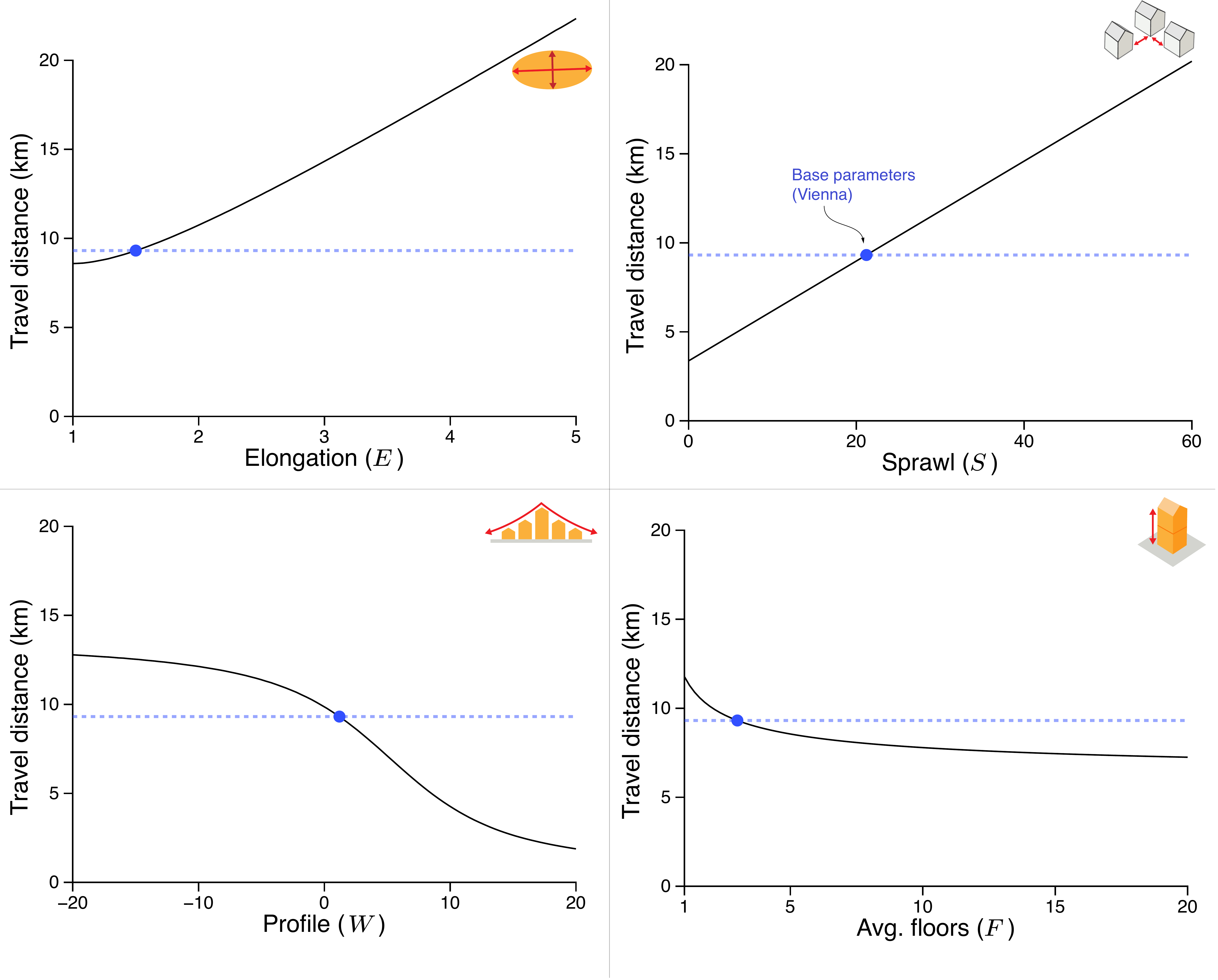}
    \caption{Effect of urban form parameters (horizontal axis) on the average commuting distance in km in the city (vertical axis). The average travel distance of the base configuration (which resembles Vienna) is shown as a blue circle.}
    \label{fig:parametersVsTravelDistance}
\end{figure}
}

{
The average travel distance increases significantly as a city becomes more elongated, even when the total urban area remains fixed. In a circular layout ($\elongation$ = 1), nodes within the street network are more centrally connected, enabling shorter, more direct routes. As elongation increases, these nodes stretch along a longer axis, forcing travel paths to extend across the grid, thereby increasing both the time and energy required for mobility. This effect is nonlinear: at $\elongation = 2$, mean travel distance increases by approximately 25\% relative to the circular baseline, while distances grow rapidly beyond this point, rising by roughly 68\% at $\elongation$ = 3 and more than doubling at large values. No saturation is observed within the tested range, however, we expect that a maximum travel distance would be reached for an extreme value of $E$ when all buildings are arranged in a single row, suggesting that continued elongation consistently degrades network efficiency and complicates the provision of uniformly accessible urban services \cite{prieto2023arguments}.
}

{
Average travel distance increases with sprawl. As more unused space is introduced between buildings, the urban footprint expands proportionally, resulting in longer commuting paths. Increasing sprawl from $\unusedArea = 0$ to $\unusedArea = 25$ \si{\m} raises the average travel distance from 3.4 \si{\km} to 4.9 \si{\km}, an increase of approximately 44\%. Unlike elongation, which deforms the city’s network topology along a dominant axis, sprawl expands the urban fabric radially by adding unused land per parcel. This form of expansion uniformly increases travel distances without altering the overall shape of the network. Importantly, no saturation effect is observed, indicating that continued sprawl consistently undermines spatial accessibility across its entire range.
}

{
Vertical growth is not merely a technical adjustment, but also a significant design decision that accommodates population growth without expanding urban areas. Assuming a constant volume, as the building height increases, the city's surface area decreases, resulting in a shorter average travel distance. Moving from expansive single-story buildings to low-rise development (e.g., $\height$ = 3 floors) while keeping the number of residents per building constant can shorten travel distances by over 10\%. At the same time, further increases to mid-rise levels yield diminishing returns. For instance, expanding from 3 to 12 floors reduces the average distance by approximately 5\%. This indicates that most of the accessibility gains from vertical expansion are realised early, with the curve flattening at higher floor counts. This diminishing return suggests that beyond a certain threshold, increasing building height is less effective as a strategy for improving urban accessibility. However, taller buildings in real cities typically house more residents. Increasing residential occupancy per building also substantially reduces travel distances, indicating that the combined effect of vertical growth and increased density may offer even greater benefits for reducing urban travel distances (details in Supplementary Materials B).
}

{
While the average number of floors influences overall density, the city's profile is an even more powerful determinant of travel efficiency. When taller buildings, and thus higher densities, are concentrated in the city centre in a pyramid profile, average travel distances drop substantially. A slight increase in the density of the city centre reduces travel lengths, and the effect intensifies nonlinearly: moderately peaked profiles ($\profile \approx 2-3$) already lower distances by 20–25\%, while extreme values in a needle city ($\profile > 15$) can reduce the average travel distance by up to 80\% compared to a uniform height distribution. This occurs because the central concentration brings most origins and destinations closer together, turning the city core into a dominant node of activity. In contrast, a bowl profile where density is shifted toward the periphery has the opposite effect. A flat city leads to gradual increases in travel, but the effect becomes pronounced with stronger profiles, as average commuting distance rises by 30–40\% in extreme ring cities ($\profile < -15$). These results suggest that increasing density near the city centre can yield substantial efficiency gains, while dispersing density outward consistently reduces accessibility with no indication of saturation.
}


\section{Discussion}

{
Many factors affect urban forms. The lack of planning, natural and political barriers, and transportation limitations have played a significant role, but policy decisions also have a substantial impact. Several factors affect most cities in similar ways (e.g., the need for accessibility, changes in transportation technologies, and an increase in living standards). While it has been argued that cities follow similar trajectories \cite{Carra2019, LEMOINERODRIGUEZ2020103949}, there are plenty of divergent examples. The vertical and compact growth of Hong Kong, the pyramid shape of Dubai, and the sprawling expansion of many cities in the US show that cities do not converge to a similar urban form. Overall, cities are subject to a range of factors that can act as pressures for either densification or sprawl, and identifying these factors and their interaction with sustainability can be important when debating policy responses.
}

{
The four key parameters of urban shape in this model strongly influence a city's average travel distance. Even small changes in factors such as elongation, building height, and profile can have a significant impact \cite{biljecki2022global}. These results highlight the importance of considering these elements when planning urban growth. While terrain, rivers, and seas may limit the planning, adjusting vertical profiles and reducing unused space can further reduce travel distances. Compared to studies that focus on more detailed aspects of urban form, such as street patterns, intersection density, and land use mix, our findings suggest that broader shape attributes have a more significant impact on travel distances and energy costs. This raises questions about traditional studies on how urban form influences travel behaviour, which often concentrate on smaller-scale factors that sometimes show conflicting or minimal effects \cite{stead2001relationships, berrill2024comparing, dalde2025effects}. In practice, larger-scale urban form may already shape or constrain travel patterns, making cross-city comparisons in those behaviour studies challenging, even when socio-economic factors are controlled for.
}

{
Our simulation of urban forms and the interaction between different parameters emphasises that urban planning is a multi-dimensional problem. Planners need to consider a range of metrics, including the total area, average travel distance, sprawl indices, and the city's profile, to achieve an optimal balance. For example, increasing the number of persons per building reduces both the total building footprint area and the average travel distance; however, if done excessively, it may lead to an overly crowded space or even slums \cite{pierce2017basic}. Thus, the optimal city shape will often be the result of iterative modelling and stakeholder engagement, where different parameters are balanced against the community's long-term vision.
}

{
Reducing urban energy consumption contributes to long-term sustainability \cite{toth2012energy}. While reducing average travel distances provides clear benefits, the urban form must also cater to human-scale experiences. Excessively high densities might compromise access to services \cite{pierce2017basic}, daylight, open space, and even community interaction if not appropriately managed. The design trade-offs here involve balancing the efficiency gains from vertical densification with the need for spaciousness, greenery, and a comfortable living environment. The simulated cities and their shapes are a simplification of a myriad of urban elements. We do not estimate congestion, which can be very important as it significantly affects the quality of life for citizens, drives decisions, and contributes to energy use and emissions. We do not consider how city form affects the feasibility of public versus private transportation; therefore, we do not differentiate our simulated commuters by travel mode. Extensions of our model could include a mode choice, allowing us to investigate the interplay between urban form, travel mode choice, congestion, and energy use. However, this is intertwined with travel modes: using a faster travel mode (e.g. cars), people might take longer commutes (considering the travel time budget hypothesis \cite{Marchetti1994, Mokhtarian2004, Kung2014}). Urban form combined with transportation infrastructure layout can have disparate effects on different travel models with respect to accessibility.
}

{
One of the challenges in planning research is effectively communicating complex findings to the public. This tool helps overcome that challenge by making these concepts easier to understand and relate to everyday life. The model, interactive simulations, and urban calculator provide valuable insights for city planners and policymakers, as well as a tool to engage the general public. It demonstrates how changes in a city's shape, achieved by adjusting individual parameters, can impact average travel distances.
}

\section{Methods}\label{sec:methods}

\subsection{Modelling Virtual Cities}\label{ssec:modelling-a-city}

{
We model cities as a street network $\network = (\vertices,\connections )$ and the heights of the buildings $\heightsOfBuildings$. 
Here, $\vertices = \{v_i, v_j, \dots\}$ represents the set of street crossing positions in a two-dimensional Euclidean space. 
The set of undirected edges between crossings is denoted by $\connections = \{\connection_i, \connection_j, \dots\}$, and the edges correspond to the streets of our city, whereas $\lVert \connection_i \rVert = \lVert v_i - v_j \rVert$ is the length of the street.
Each building, its size, shape, and location are given by the four streets 
$c_i^i, c_j^i, c^i_k\text{, and } c_l^i$ 
that surround the property. 
In the modelled cities, each building is surrounded by four streets, and each crossing is part of a maximum of four properties.
Hence, $\network$ is a quad mesh, where each property is a face, each street crossing is a vertex, and each street is an edge. 
}

{
We model the shape of a city using a set of parameters, which include the elongation $\elongation$, the sprawl as $\unusedArea$, the average height of buildings $\height$, and the city's profile $\profile$. 
}

{
We model the shape of the overall footprint as an ellipse enclosing $\network$. The ratio of the longest and the shortest extension of the ellipse equals $\elongation$. Hence, a city with $\elongation = 1$ has a circular shape, and a city with $\elongation = 3$ has a stretched elliptical layout (Fig. \ref{fig:city-varibles}).
}

{
The sprawl $\unusedArea$ describes the interbuilding distance between two neighbouring buildings and is given as $\unusedArea = \lVert \connection \rVert - \sqrt{a}$. In our model, the sprawl $\unusedArea$ describes every space not occupied by buildings, such as streets, green spaces, sidewalks, or parking lots, which can also be understood as open space \cite{ewing1997angelesstyle}.
}

{
We consider the height of buildings in a city to be a function that depends on the distance to the city centre. For some location at distance $d$, we vary the height $h(\profile, d)$ according to the profile parameter $\profile$ as 
\begin{equation}\label{eq:profile}
    h(\profile, d) = s(d + 1)^{-\profile }, 
\end{equation}
where $s$ is a scale value to ensure that the volume (equalling the apartment space) of the city stays constant when $\profile$ is altered. It is given as the ratio between the rotational volumes of $h(0,d)$ and $h(\profile, d)$ (details in Supplementary Materials A). The idea is that it is possible to change the profile of the city, keeping the city's volume fixed (so independent of the different values of $\profile$). The distance to the centre $d$ in the parameter space of the ellipse that encloses the city, with $d = 0$ in the centre and $d = 1$ at the ellipse's border.
This distance can be calculated by $d = \sqrt{ (v_x / m)^2 + (v_y / M)^2 }$, while $v$ is the delta vector between the centre of the building and the centre of the ellipse, and $v_x$ and $v_y$ its components x and y, respectively.
}


{
The average number of floors of buildings is written as $\height$. In our model, $\height$ does not affect the volume of buildings and the required number of buildings to house the defined population. If $\height$ increases, the average number of floors of buildings increases, but at the same time, the footprint of the buildings decreases to preserve the apartment space of the individual buildings.
}

{
We illustrate the effect of each parameter on the inner-city travel distance, measured by the average distance a person has to travel to reach from one point to another arbitrary point in the city. To compute this for one virtual city, we sample 100,000 pairs of positions and calculate the average distance between them. Repeating this process while altering a single city parameter allows us to observe the effect of that parameter on the travel distance. 
}

\subsection{Computing the urban morphology}\label{ssec:computing-morphology}
{
Based on the parameters $\elongation$, $\unusedArea$, $\height$ and $\profile$, the population $\inhabitants$, the average number of persons per building $\perBuilding$, and the floor area per resident $\perresident$, we compute a cities network $\network$ and the corresponding heights of the buildings $\heightsOfBuildings$:  
\begin{equation}
    \network, \heightsOfBuildings = g(\elongation,\unusedArea,\height,\profile,\inhabitants,\perBuilding, \perresident).
\end{equation} 
}

{
First, we define an ellipse that encloses the whole city. The size of the ellipse depends on $\unusedArea$ and the footprint of the individual buildings $a$ given as $a = \perresident \perBuilding / \height$. Based on $a$ and the number of buildings $b = \lceil \perBuilding / \inhabitants \rceil$, we can compute the ellipse characterised by its minor extension $m$ and major extensions $M$ as 
\begin{equation}
     m = 2  \frac{\unusedArea}{\elongation} \sqrt{\frac{a b}{\pi}}, 
\end{equation}
}
and
\begin{equation}
    M = \elongation^2 m.
\end{equation}
{
We limit the city's layout to equal long $\connection$ with a length of $\lVert \connection \rVert = \lVert v_j - v_i \rVert = \unusedArea \sqrt{a}$. Following this, we create a rectangular grid that is aligned with the ellipse with $\lceil m / \lVert \connection \rVert \rceil$ rows and $\lceil M / \lVert \connection \rVert \rceil$ columns. Then, we remove the grid cell with the longest parametric distance $d$ to the ellipse's centre, so the number of grid cells equals the number of buildings $b$. The number of floors of an individual building $ h_i \in \heightsOfBuildings$ corresponds with the likelihood that it is selected as a start or end point for the path simulation. This is computed by $h(W, d_i)$, whereas $d_i$ corresponds to the distance of the building to the city centre. 
}

{
To compute the average travel distance of a city, a proxy for travel-related energy consumption \cite{prieto2023scaling}, we sample 100,000 journeys in $\network$. Each journey has a randomly selected starting and endpoint $\vertex_i, \vertex_j \in \vertices$. The likelihood that a street crossing $\vertex_i$ is selected is proportional to the average number of floors of all neighbouring buildings. Hence, our city has a regular grid, and the average Manhattan distances between the start and endpoints give the average travel distance. The resulting average travel distance differs from the average length of all possible shortest paths because not all $\vertex$ are equally likely to be picked for a city with $\profile \neq 0$.
}

{
We have made the source code of our model implementation publicly available at \url{https://github.com/TobiasBat/Cities-Morphology}. The repository includes all necessary resources to reproduce the experiments outlined in this paper, our implementation of the parametric city model, and the source code of the visualisation page, Cities Morphology. 
}

\section{Supplementary materials}

\renewcommand{\figurename}{Supplementary Figure}

\subsection*{A - Impact of additional parameters on travel distances} 

{
To isolate the effects of specific urban shapes, we vary the shape parameters while holding other factors constant. In addition to the core city shape characteristics discussed in the main text (sprawl $\unusedArea$, elongation $\elongation$, average building height $\height$, and profile $\profile$), we analyse three further parameters that impact travelling distances: total population (\inhabitants), area per resident ($\perresident$), and residents per building ($\perBuilding$). While these variables are not strictly city-shape parameters in the sense of spatial form, they influence the density and scale of the urban system and provide important context for understanding urban accessibility (Supplementary Figure  \ref{fig:parametersVsTravelDistanceSubMaterial}).

\begin{figure}[h!]
    \centering
    \includegraphics[width=\linewidth]{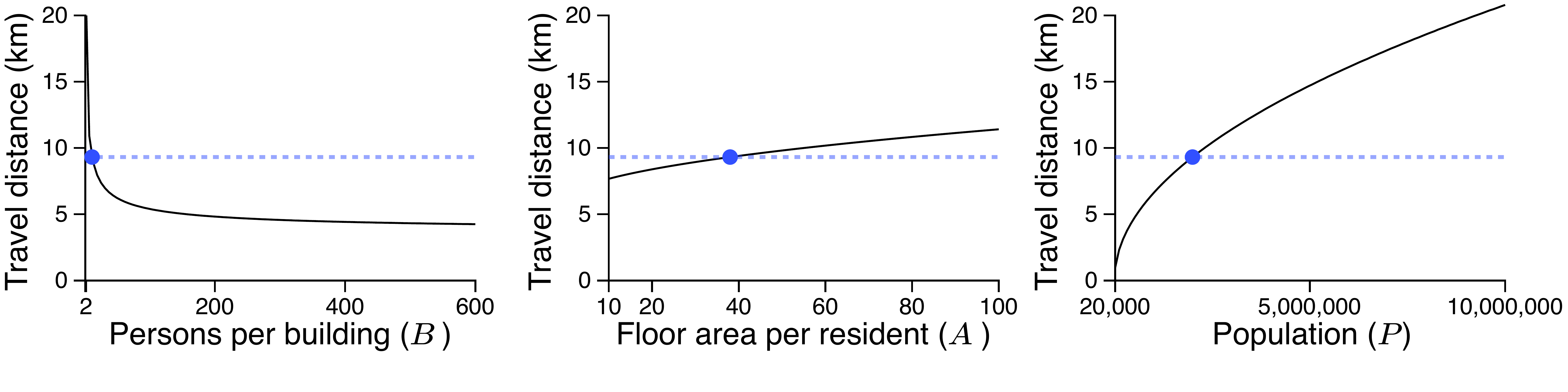}
    \caption{Impact of other urban indicators in terms of average travel distance.}
    \label{fig:parametersVsTravelDistanceSubMaterial}
\end{figure}
}

{
Average travel distance increases with the total population. However, this increase is sublinear. For example, doubling the population from 1 to 2 million increases the average travel distance by only about 40\%. A similar effect is observed when expanding from 5 to 10 million residents. This square-root-like scaling implies that even substantial growth in population or area leads to only gradual increases in commuting distances. The most significant jumps in average travel occur when transitioning from a small town to a mid-sized city; beyond that, additional growth adds relatively little to travel requirements. This sublinear scaling phenomenon is common in spatial systems. For example, by looking at the distribution of nearly 200 million buildings in African cities, the urban footprint increases roughly in proportion to the number of residents times the area allocated per person, the average inter-residential distance (which for a uniform distribution scales as the square root of area) exhibits this square-root relationship \cite{prieto2023scaling}. These diminishing returns offer a potential advantage in urban planning: cities can expand in size without incurring proportionally longer commutes, especially when density and spatial organisation are optimised.
}

{
Varying the land area allocated to each resident yields a near-linear relationship with travel distance. Increasing area per resident reduces overall population density and leads to longer average commutes. Over the tested range, average travel distance increased steadily by 16\%. While this effect is less pronounced than that observed for sprawl or profile, it reinforces the intuitive principle that more dispersed populations necessitate longer travel paths. Notably, this relationship showed no clear threshold or inflexion point: the increase in travel distance was gradual and consistent, without evidence of saturation. Thus, while increasing personal space may provide certain livability benefits, it comes at a measurable cost in accessibility.
}

{
Finally, increasing the number of residents per building has a substantial effect on travel distance, particularly when moving from very low to moderate occupancy levels. At the lowest end of the spectrum ($\perBuilding$ = 2 residents per building), average travel distances are among the highest observed in the simulations. As building occupancy increased to 50 residents, travel distances decreased by 75\%. This substantial gain highlights the efficiency of clustering residents into shared structures. However, the benefit of growing occupancy tapers off beyond this point. Increasing the number of persons per building to 200 or more yielded only marginal additional improvements, with the average travel distance approaching an asymptotic minimum of approximately 3.6 km. Under a fixed volume assumption, this pattern indicates diminishing returns to building-level density: while initial increases in occupancy significantly compress the urban footprint and improve accessibility, the gains level out as buildings become highly populated.
}

\subsection*{B - Keeping the volume of a city fixed}

{
The profile of a city is given by the height of buildings $h(\profile, x)$ as the distance to the city centre $x$ changes according to
\begin{equation}\label{eq:profile}
    h(\profile, x) = s(x + 1)^{-\profile }.
\end{equation} 
To preserve the constructed volume of a city (and the average number of floors $\height$) fixed when altering $\profile$, we multiply the heights of the buildings by a scale value $s$ depending on the profile of the city $\profile$. The value is given as the ratio of the rotational volume ($V$) of the city with $\profile = 0$ and the rotational volume of the city $V_\profile$ written as 
 $s = V_0/V_\profile$, with 
\begin{eqnarray*}
    V_\profile &=& 2 \pi  \int_0^1 x (x+1)^{-\profile} \, dx,
\end{eqnarray*}
which we compute numerically to obtain a fixed constructed surface. The distance $x$ is given in the parameter space of the ellipse that encloses the city, ranging from 0 to 1 in all directions. Therefore, $h(\profile,x)$ is independent of the sprawl $\unusedArea$ and the elongation $\elongation$ of a city and since they affect the Euclidean distance of a building to the city centre, rather than the ellipse's parameter space.
}

\subsection*{C - Interactive tools}

\begin{figure}[h!]
    \centering
    \includegraphics[width=11 cm]{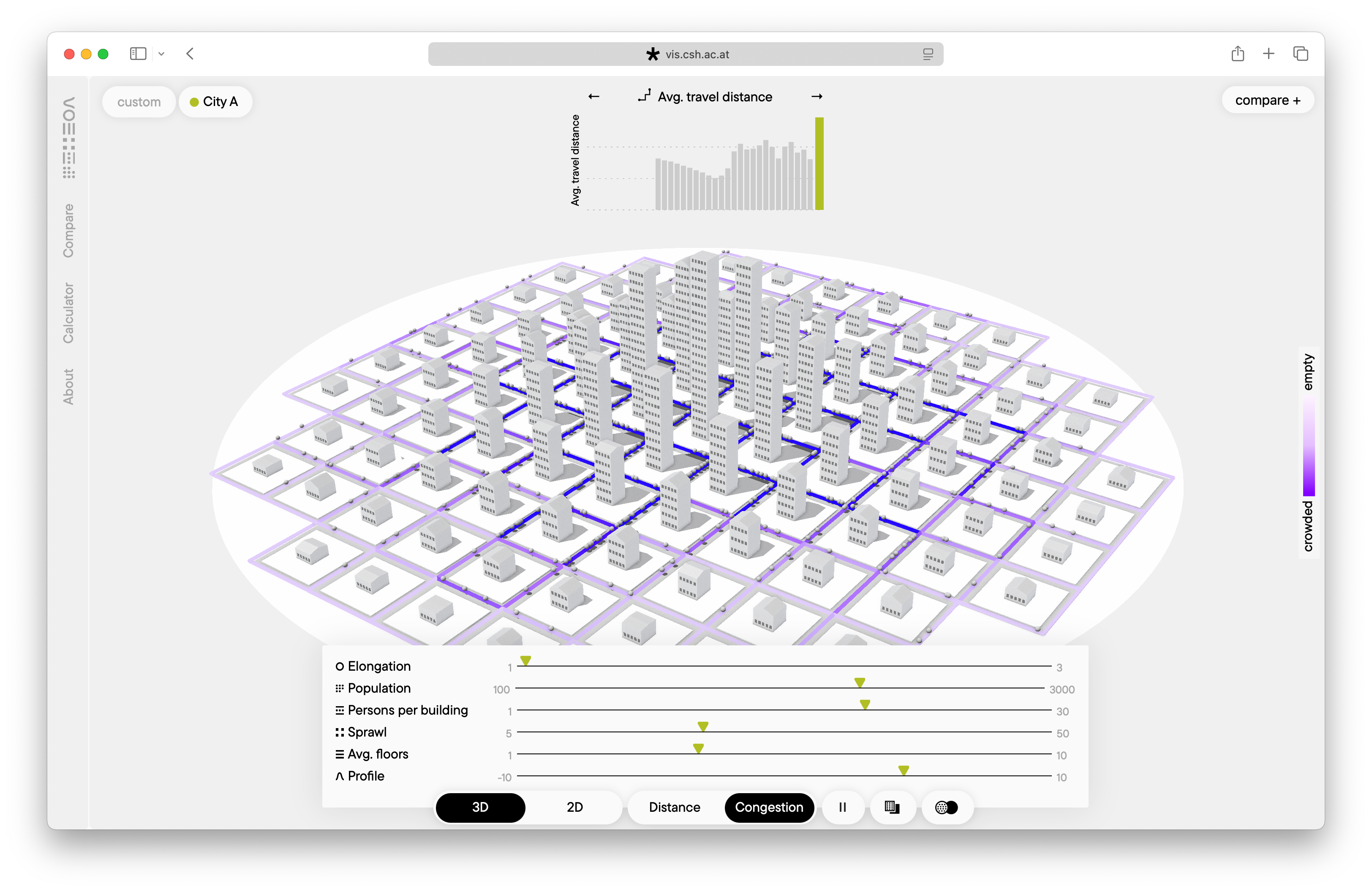}
    \caption{Simulation of different city layouts, enabling six parameters to be changed, and to obtain the average commuting distance, as well as the distribution of the congestion. Tool available at \url{https://vis.csh.ac.at/cities-morphology}}
    \label{fig:simulator}
\end{figure}

We developed an interactive simulator that allows users to modify key parameters defining a city’s profile and immediately observe the impact on travel distances (Supplementary Figure \ref{fig:simulator}). The system computes the average travel distance between any two individuals within the city under varying forms and densities.

\begin{figure}[h!]
    \centering
    \includegraphics[width=11cm]{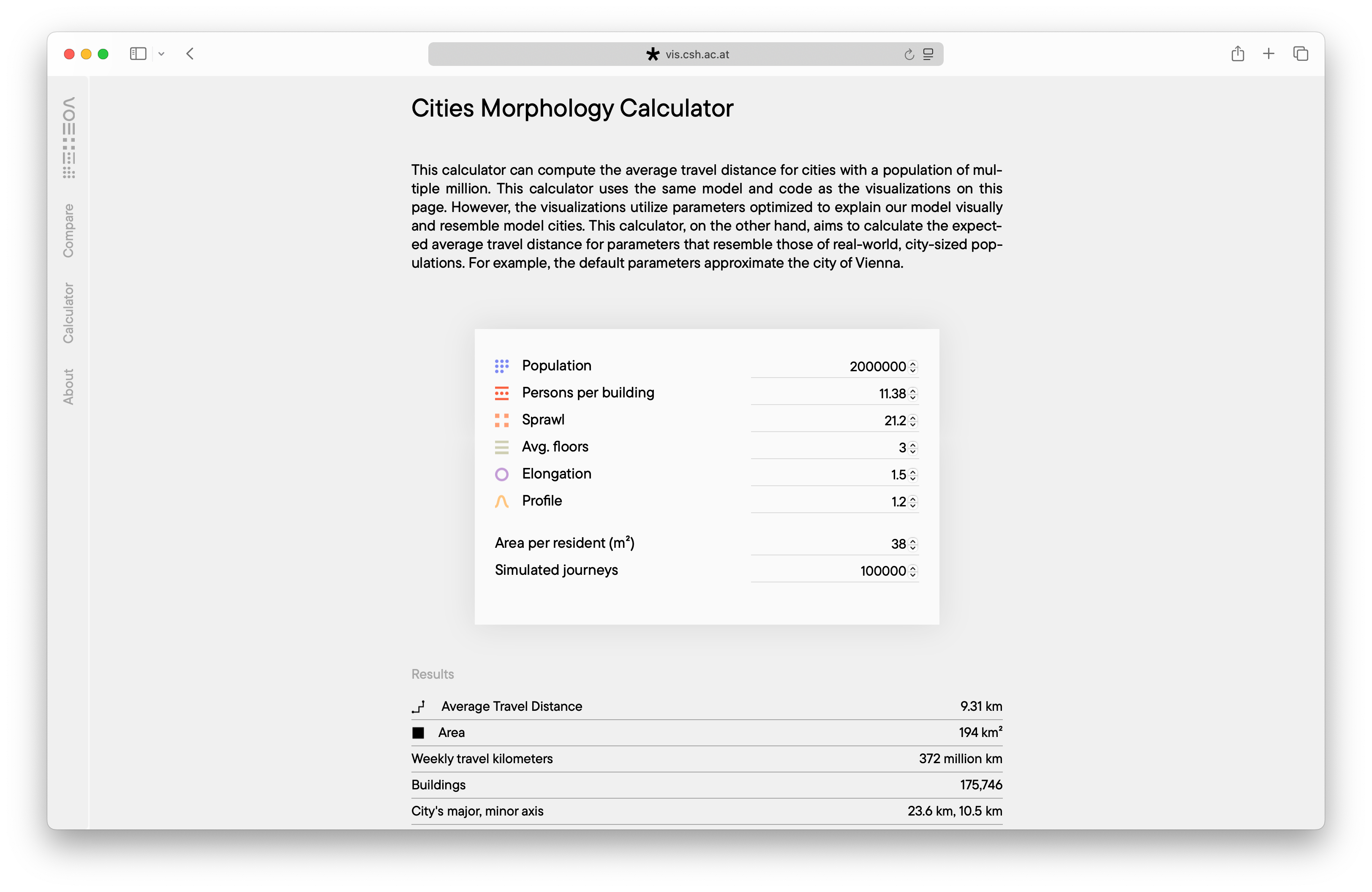}
    \caption{To quantify the impact of different layouts, we added a calculator which runs thousands of simulations and outputs to the user the average distance travelled in that city.}
    \label{fig:calculator}
\end{figure}

We also developed a calculator that enabled us to simulate thousands of urban layouts and detect, in terms of commuting distance, the impact of different urban forms (Supplementary Figure \ref{fig:calculator}). The process in the background is also running thousands of simulations and computing the average distance between them.

\end{document}